\documentstyle[11pt]{article}

\let\sll=\l                     % slashed (suppressed) l (Polish)
\def\vu{\varepsilon}

\def\d{\delta}
\def\e{\epsilon}                % Also, \varepsilon
\def\h{\eta}

\def\j{\psi}
\def\l{\lambda}

\def\o{\omega}
\def\q{\theta}                  %       \vartheta
\def\t{\tau}

\def\F{\Phi}
\def\G{\Gamma}

\newcommand{\extraspace}{\addtolength{\abovedisplayskip}{2mm}
                        \addtolength{\belowdisplayskip}{2mm}
                        \addtolength{\abovedisplayshortskip}{2mm}
                        \addtolength{\belowdisplayshortskip}{2mm}}
\newcommand{\be}{\begin{equation}\extraspace}

\newcommand{\ee}{\end{equation}}

\newcommand{\bea}{\begin{eqnarray}\extraspace}
\newcommand{\beastar}{\begin{eqnarray*}\extraspace}
\newcommand{\eea}{\end{eqnarray}}
\newcommand{\eeastar}{\end{eqnarray*}}

\newcommand{\nonu}{\nonumber \\[2mm]}

\newcommand{\strutje}{\rule[-1.5mm]{0mm}{5mm}}

\newcommand{\half}{\frac{1}{2}}

\newcommand{\ind}{{\rm ind}}

\newcommand{\del}{\partial}

\newcommand{\bdel}{\bar{\partial}}
\newcommand{\nablab}{\overline{\nabla}}

\def\ttt{\tilde{t}}
\def\tg{\tilde{g}}
\def\tu{\tilde{u}}

\def\tc{\tilde{c}}
\def\tk{\tilde{k}}

\def\tG{\widetilde{G}}
\def\tGa{\widetilde{\Gamma}}
\def\tU{\widetilde{U}}
\def\tT{\widetilde{T}}
\def\tW{\widetilde{W}}

\def\tzeta{\tilde{\zeta}}
\def\tgamma{\tilde{\gamma}}
\def\tkappa{\tilde{\kappa}}
\newcommand{\dd}[1]{\frac{\d}{\d #1}}
\newcommand{\DD}[2]{\frac{\d #1}{\d #2}}

\newcommand{\np}{Nucl.\ Phys.\ }

\newcommand{\pl}{Phys.\ Lett.\ }

\catcode`\@=11
\def\@afterindentfalse{\let\if@afterindent=\iftrue}
\catcode`\@=12

\begin{document}

\font  \biggbold=cmbx10 scaled\magstep2
\font  \bigbold=cmbx10 at 12.5pt
\font  \bigreg=cmr10 at 12pt

\baselineskip = 15pt

\noindent March 1993 \hfill LBL-323778, UCB-PTH-93/07\\
$\strutje$ \hfill  KUL-TF-93/10\\
$\strutje$ \hfill  hep-th/9304020\\

\vspace{4mm}

\begin{center}
{\large THE RELATION BETWEEN LINEAR AND NON-LINEAR}
{\large $N=3,4$ SUPERGRAVITY THEORIES
\footnote{This work was supported in part by the Director,
Office of Energy Research, Office of High Energy and Nuclear Physics,
Division of High Energy Physics of the U.S. Department of Energy
under Contract DE-AC03-76SF00098 and in part by the National Science
Foundation under grant PHY90-21139.} }\\

\vspace{1cm}

{\bf \centerline{Alexander Sevrin${}^1$, Kris Thielemans${}^2$ and
Walter Troost\footnote{Bevoegdverklaard Navorser NFWO,Belgium}${}^{2}$}}
 \vskip .3cm
{\baselineskip = 12pt
\centerline{\sl{1. Department of Physics}}
\centerline{\sl{University of California at Berkeley}}
\centerline{\sl{and}}
\centerline{\sl{Theoretical Physics Group}}
\centerline{\sl{Lawrence Berkeley Laboratory}}
\centerline{\sl{Berkeley, CA 94720, U.S.A.}}}
\vskip .2cm
{\baselineskip = 12pt
\centerline{\sl{2. Instituut voor Theoretische Fysica}}
\centerline{\sl{Universiteit Leuven}}
\centerline{\sl{Celestijnenlaan 200D, B-3001 Leuven, Belgium}}}
\end{center}

\vskip 1.4cm
\centerline{\bf Abstract}
\vskip 0.5cm
{\baselineskip=14pt
The effective actions for $d=2$, $N=3,4$ chiral supergravities
with a linear and a non-linear gauge algebra are related to each other
by a quantum reduction, the latter is obtained from the former by
putting quantum currents equal to zero. This implies that the
renormalisation factors for the quantum actions are identical.
}

\newpage

\pagestyle{plain}
\renewcommand{\thepage}{\roman{page}}
\setcounter{page}{2}
\mbox{ }

\vskip 1in

\begin{center}
{\bf Disclaimer}
\end{center}

\vskip .2in

\begin{scriptsize}
\begin{quotation}
This document was prepared as an account of work sponsored by the United
States Government.  Neither the United States Government nor any agency
thereof, nor The Regents of the University of California, nor any of their
employees, makes any warranty, express or implied, or assumes any legal
liability or responsibility for the accuracy, completeness, or usefulness
of any information, apparatus, product, or process disclosed, or represents
that its use would not infringe privately owned rights.  Reference herein
to any specific commercial products process, or service by its trade name,
trademark, manufacturer, or otherwise, does not necessarily constitute or
imply its endorsement, recommendation, or favoring by the United States
Government or any agency thereof, or The Regents of the University of
California.  The views and opinions of authors expressed herein do not
necessarily state or reflect those of the United States Government or any
agency thereof of The Regents of the University of California and shall
not be used for advertising or product endorsement purposes.
\end{quotation}
\end{scriptsize}

\vskip 2in

\begin{center}
\begin{small}
{\it Lawrence Berkeley Laboratory is an equal opportunity employer.}
\end{small}
\end{center}

\newpage
\renewcommand{\thepage}{\arabic{page}}
\setcounter{page}{1}

\baselineskip=18pt

\section{Introduction}
\setcounter{equation}{0}
When a Lie algebra is generalised to a commutator algebra that is not
linear in its generators, but contains quadratic or higher order
polynomials as well, one obtains a non-linear algebra. Especially the
infinite-dimensional variety showing up in CFT have recently been studied
intensively, the most celebrated class being the W-algebras
(for a review, see \cite{pbks}), and in particular the W$_3$-algebra
\cite{zamo}.

Among the properties that the $W_3$-algebra, for one, shares with linear
algebras, is a remarkable renormalisation property of the quantum theory
that arises when one couples the currents $J$ generating these algebras to
gauge fields $A$. The resulting induced action for the gauge fields is
non-zero due to anomalies (central terms, quantum corrections) in the
current commutators as compared to the classical Poisson brackets, and one
can take this induced action as a starting point to quantise the gauge
fields. For linear current algebras like affine Lie algebras and the
Virasoro algebra this induced action $\Gamma $ is proportional to the
central charge, $\Gamma _{ind}[A] = c \Gamma ^{(0)} [A]$. Also, the
effective action $\Gamma _{eff} [A]$, or equivalently $W$, the generator
of connected Green functions for $A$\footnote{Hereafter also called
'effective action' for brevity. The symbol used should resolve possible
doubts on which functional is meant.}, is related \cite{polyakov,KPZ}
to the same basic functional by a field- and coupling
renormalisation $\Gamma _{eff}[A]= Z_\Gamma  \Gamma ^{(0)} [Z_AA]$. There
are several methods to compute these $Z$-factors \cite{KPZ,Zfac methods}, with
general agreement for $Z_\Gamma $ and varying proposals for $Z_A$ - for a
discussion see \cite{ruud}. For non-linear algebras on the other hand, the
dependence of the induced action on the central charge is not simply
proportionality, but instead it can be expanded in powers of $1/c$, $
\Gamma _{ind} [A] = \displaystyle{\sum_{i\geq0}} c^{1-i} \Gamma ^{(i)}
(A)$. It is remarkable that nevertheless, for the quantum theory based on
this action, the renormalisation property still holds: the effective action
is still equal, up to renormalisation factors, to the 'classical' ({\it
i.e.} lowest order in $c$ ) term $\Gamma ^{(0)}[A]$ of the induced action.
This was shown for $W_3$ to first order (and conjectured to be true to all
orders) in \cite{ssvnb}.  This was recently proved in \cite{dbg}, and
extended in \cite{zfactors} to arbitrary extensions of the Virasoro algebra
that can be obtained from a Drinfeld-Sokolov reduction \cite{drisok} of
WZW models.

In this note we point out that there are a few cases, {\it viz.} $N=3,4$
supergravities where the renormalisation of the linear and non-linear
effective actions is intimately related, due to the simple relation that
exists between the $N=3,4$ linear \cite{adem,n4} and non-linear
\cite{bershkniz,goddard} superconformal algebras.  Namely, we will show in
both cases that the effective action $W$ of the non-linear theory results
from that of the linear theory by putting to zero an appropriate set of
currents (or integrating out an appropriate set of fields for $\Gamma $).
By the same token, we will then have shown that for $N=3,4$ chiral
supergravity the same type of cancellations occur, as referred to above for
$W_3$.  Namely, non-leading terms in the central charge in $\Gamma
_{ind}[A]$ cancel with quantum contributions to $\Gamma_{eff}$.
The identity of the renormalisation factors also follows.
\section{$N=3$ Supergravity}
\setcounter{equation}{0}

Both $N=3$ superconformal algebras contain the
energy-momentum tensor $T$, 3 supercharges $G^a$, $a\in\{1,2,3\}$ and an
$so(3)$ affine Lie algebra, $U^a$, $a\in\{1,2,3\}$. The linear one
\cite{adem} contains in addition a dimension $\half$ fermion $Q$.
The operator product expansions
(OPEs) of the generators are (we use tildes for the non-linear
algebra) :
\be
\begin{array}{rclcrcl}
T   \,   T    &=& \frac{c}{2} [1] &\hspace*{1cm}&
\tT \,   \tT  &=& \frac{\tc}{2} [1] \nonu
T   \,   \F   &=& h_{\F} [\F] &&
\tT \,   \tilde{\F}  &=& h_{\F} [\tilde{\F}] \nonu
G^a \,   G^b  &=& \d^{ab}\frac{2c}{3} [1] -\vu^{abc} 2 [U^c] &&
\tG^a\,  \tG^b&=& \d^{ab}\frac{2(\tc-1)}{3} [1]
                  -\frac{2(\tc-1)}{\tc+1/2}\vu^{abc} [\tU^c]\nonu
&&&&&&            +\frac 3 {\tc+1/2} [\tU^{(a} \tU^{b)} -
                  { 2\tc+1 \over 3\tc} \delta^{ab} \tT ] \nonu
U^a \,   U^b  &=& -\frac {c}{3} \d^{ab} [1] +\vu^{abc} [U^c] &&
\tU^a \, \tU^b&=& -\frac {\tc+1/2}{3} \d^{ab} [1] +\vu^{abc} [\tU^c] \nonu
U^a \,   G^b  &=& \d^{ab} [Q]+ \vu^{abc}[G^c] &&
\tU^a\,  \tG^b&=& \vu^{abc} [ \tG^c ] \nonu
Q   \,   G^a  &=& [U^a] &&&& \nonu
Q   \,   Q    &=& -\frac c 3[1], &&&&
\end{array} \label{eq:algs N=3}
\ee
where $ h_{\F}=h_{\tilde\F}=\frac 3 2 ,\ 1,\ \frac 1 2$ for $\F=G^a,\
U^a,\ Q$.

The relation between the two algebras is \cite{goddard} that $Q$ commutes
with the combinations that constitute the non-linear algebra
\bea
\tT&\equiv&T-\frac{3}{2c}Q\del Q,
\nonu
\tG^a&\equiv&G^a+\frac 3 c U^aQ,
\nonu
\tU^a&\equiv&U^a,\label{decoup}
\eea
while the central charges are related by $\tc = c-1/2$.

The induced action $\G$ is defined by
\bea
\lefteqn{Z[h,\j,A,\h ] = \exp \Big[ -\G [h,\j,A,\h ]\Big] =} \nonu
&&\Big\langle\exp\Big[ -\frac{1}{\pi}
\int d^2\,x \Bigl( h(x) T(x) + \j_a(x) G^a(x)
 + A_a(x) U^a(x) + \h (x) Q(x)\Bigr)\Big]
\Big\rangle.
\label{eq:indactL3}
\eea
and similarly, without the $\h$-field, for the non-linear induced
 action $\tilde \G$
These actions are completely determined by considering their transformation
 properties under $N=3$ supergravity transformations.
These transformations read, for the linear case:
\bea
\d h&=&\bdel \e + \e \del h - \del \e h + 2 \q^a \j_a,
\nonu
\d \j^a&=&\bdel \q^a + \e \del \j^a - \frac 1 2 \del \e \j^a+\frac 1 2 \q^a\del
h-\del \q^a h
 -\vu^{abc}(\q^bA^c+\o^b\j^c),
\nonu
\d A^a&=&\bdel \o^a+\e \del A^a-\vu^{abc}(\del \q^b\j^c-\q^b\del \j^c)
+\q^a\h-\vu^{abc}\o^bA^c-\del\o^ah+\t\j^a,
\nonu
\d \h&=&\bdel \t + \e \del \h +\frac 1 2  \del \e \h +\q^a\del
A_a-\del\o^a\j_a
 -\frac 1 2 \t\del h -\del \t h.\label{eq:sutranslin3}
\eea
For the non-linear case, they are the same, except that
 there is of course no field $\h$ and no parameter $\t$, and
$\d A_a$ contains a $\tc$ dependent extra term
\be
\d A^a_{\mbox{\footnotesize extra}} =
 \frac{3}{2\tilde c}\vu^{abc}(\del \q^b\j^c-\q^b\del \j^c).
 \label{deltaAextra}
\ee
The anomaly for the linear theory is:
\be
\d  \G [h,\j,A,\h
]=-\frac{c}{12\pi}\int\e\del^3h-\frac{c}{3\pi}\int\q^a\del^2\j_a +
\frac{c}{3\pi}\int\o^a\del A_a+\frac{c}{3\pi}\int\t\h.\label{ano1}
\ee
Defining\footnote{All functional derivatives are left derivatives.}
\be
t=\frac{12\pi}{c}\frac{\d\G }{\d h} \hspace{1cm}
g^a=\frac{3\pi}{c}\frac{\d\G }{\d \j^a} \hspace{1cm}
u^a=-\frac{3\pi}{c}\frac{\d\G }{\d A^a} \hspace{1cm}
q=-\frac{3\pi}{c}\frac{\d\G }{\d \h}
\ee
we obtain the Ward identities for the linear theory by combining eqs.
(\ref{eq:sutranslin3}) and (\ref{ano1}):
\bea
\del^3h&=&\nablab t-\left(2\j_a\del + 6 \del\j_a\right)g^a
+4 \del A_au^a-\left(2\h\del-2\del\h\right)q,
\nonu
\del^2 \j_a&=&\nablab g^a-\frac 1 2 \j^at+\vu^{abc}A_bg^c+\h u^a
+\vu^{abc}\left(2\del\j_b+\j_b\del\right)u^c+\del A_a q,
\nonu
\del A_a&=&\nablab u^a-\vu^{abc}\j_bg^c+\vu^{abc}A_bu^c
-\left(\j_a\del+\del\j_a\right) q,
\nonu
\h &=&\nablab q-\j_au^a,\label{eq:WIL3}
\eea
where
\be
\nablab\F=\left(\bdel-h\del-h_{\F}(\del h)\right)\F,
\ee
with $ h_{\F}=2,\ \frac 3 2 ,\ 1,\ \frac 1 2$ for $ \F=t,\ g^a,\ u^a,\ q$.

The Ward identities provide us with a set of functional differential equations
for the induced action. Since these have no explicit dependence on
$c$, the induced action can be written as
\be
\G [h,\j,A,\h ]=c\ \G^{(0)} [h,\j,A,\h ],\label{ohwell}
\ee
where $\G^{(0)}$ is $c$-independent.

The non-linear theory can be treated in a parallel way. The anomaly is now
\bea
\d  \tGa [h,\j,A ]&=&-\frac{\tc} {12\pi}\int\e\del^3h-
\frac{\tc-1}{3\pi}\int\q^a\del^2\j_a +
\frac{\tc+1/2}{3\pi}\int\o^a\del A_a\nonu
&&-\frac{3}{\pi (\tc+1/2)}\int\q_a\j_b\left(U^{(a}U^{b)}\right)_{\mbox{eff}}.
\label{eq:anomNL2pre}
\eea
The last term, which is due to the non-linear term in the algebra eq.
(\ref{eq:algs N=3}), can further be rewritten as
\bea
\left(U^{(a}U^{b)}\right)_{\mbox{eff}}(x)&=&
\Big\langle \,\tU^{(a} \tU^{b)}(x) \exp\Big[
-\frac{1}{\pi} \int  \Bigl( h \tT + \j_a \tG^a
 + A_a \tU^a\Bigr)\Big]
\Big\rangle/\exp \Big[-\tGa\Big] \\ \label{effreg}
&=&\left(\frac {\tc+1/2}{3} \right)^2 u^a(x)\,u^b(x)\nonu
&&\qquad+\frac{(\tc+1/2)\pi}{6}
\lim_{y\rightarrow x}\biggl(\frac{\del u^a (x)}{\del A_b(y)}
 -\frac{\del}{\bdel}\d^{(2)}(x-y)\d^{ab} + a\rightleftharpoons
b \biggr). \nonumber
\eea
The limit in the last term of eq. (\ref{effreg}) reflects the point-splitting
regularization of the composite terms in the $\tG \tG$ OPE (\ref{eq:algs
N=3}).  One notices that in the limit $\tc\rightarrow\infty$, $u$ becomes
$\tc$ independent and one has simply
\be
\lim_{\tc\rightarrow\infty}
     \left( \frac {3}{\tc+1/2} \right)^2
       \left(U^{(a}U^{b)}\right)_{\mbox{eff}}(x)= u^a(x) \, u^b(x).
\label{effreglc}
\ee
Using eq. (\ref{effreg}), we find that eq. (\ref{eq:anomNL2pre}) can be
rewritten as:
\bea
\d  \tGa [h,\j,A ]&=&-\frac{\tc} {12\pi}\int\e\del^3h-
\frac{\tc-1}{3\pi}\int\q^a\del^2\j_a + \frac{\tc+1/2}{3\pi}\int\o^a\del A_a
-\frac{ \tc+1/2}{3\pi}\int\q_a\j_b u^a u^b \nonu
&&-\lim_{y\rightarrow x} \int \q^{\,(a}\j^{b)} \left(\frac{\del u^a (x)}{\del
A_b(y)}-\frac{\del}{\bdel}\d^{(2)}(x-y)\d_{ab}\right), \label{eq:anomNL3}
\eea
where the last term disappears in the large $\tc$ limit.  The term
proportional to $\int\q_a\j_b u^a u^b$ in eq.  (\ref{eq:anomNL3}) can be
absorbed by adding a field dependent term in the transformation rule for
$A$:
\be
\d^{\rm nl}_{\mbox{\footnotesize extra}}A_a=-\q_{a}\j_{b}u^b.
\ee
Doing this, we find that in the large $\tc$ limit, the anomaly reduces to
the minimal one.

Combining the non-linear transformations with eq. (\ref{eq:anomNL3}),
and defining
\be
\ttt=\frac{12\pi}{\tc}\frac{\d\tGa}{\d h}\hspace{1cm}
\tg^a=\frac{3\pi}{\tc-1}\frac{\d\tGa}{\d \j_a}\hspace{1cm}
\tu^a=-\frac{3\pi}{\tc+1/2}\frac{\d\tGa}{\d A_a},
\label{defcurr}
\ee
we find the Ward identities for $\tGa [h,\j,A ]$ (they can also
be found in \cite{gustav}):
\bea
\del^3h&=&\nablab\ttt-\left(1-\frac{1}{\tc}\right)\left(2\j_a\del +
        6 \del\j^a\right)\tg^a
+4 \left( 1+\frac{1}{2\tc}\right)\del A_a \tu^a,
\nonu
\del^2 \j_a&=&\nablab\tg^a-\left(\frac 1 2 + \frac{1}{2\tc-2}\right)
\j_a\ttt+\vu^{abc}A_b\tg^c
+ \vu^{abc} \left(2\del\j_b+\j_b\del\right) \tu^c\nonu
&&- \left(1+\frac{3}{2\tc-2}\right)\left(3\over \tc+1/2\right)^2
\j_b \left(U^{(a}U^{b)}\right)_{\mbox{eff}},
\nonu
\del A_a&=&\nablab\tu^a-\left(1-\frac{3}{2\tc+1}\right)
\vu^{abc}\j_b\tg^c+\vu^{abc}A_b\tu^c,\label{eq:WINL3}
\eea

The normalisation of the currents has been chosen so that the anomalous terms
on the l.h.s have coefficient unity.
The explicit $\tc$ dependence of the Ward identities arises from several
sources: the fact that in the non-linear algebra, eq.(\ref{eq:algs N=3})
 some couplings are explicitly $\tc$-dependent, the $\tc$ dependence of the
 transformation, eq.(\ref{deltaAextra}), and the field-non-linearity.
The dependence implies that the induced action is given by a $1/\tc$
 expansion:
\be
\tGa_\ind[h,\j,A]=
 \sum_{i\geq 0}\tc^{1-i}\tGa^{(i)}[h,\j,A] \ {}.
\ee
This is familiar from $W_3$ \cite{ssvnb}.

Turning back to the Ward identities for the linear theory eq. (\ref{eq:WIL3}),
we observe a remarkable relation with the non-linear ones. If we
take $\tilde{c}=c+1/2$ and put $q=0$, we find from the last identity in
eq.  (\ref{eq:WIL3}) that $\h =-\j_au^a$.  Substituting this back into the
first three identities in eq.(\ref{eq:WIL3}), yields precisely the Ward
identities for the non-linear theory eq.  (\ref{eq:WINL3}) in the
$c\rightarrow\infty$ limit.  Also, the extra term in the non-linear $\d
A_a$ ( eq. \ref{deltaAextra}), that was added to bring the anomaly to a
minimal form, now effectively
re-inserts the $\q^a \h$ term that dropped out of the linear
transformation, eq.(\ref{eq:sutranslin3}).  This strongly suggests that the
relation between the effective theories should be obtained by putting the
current $q$ equal to zero on the quantum level.  We will now derive this
fact.

First we rewrite eq. (\ref{eq:indactL3}) using eq.
(\ref{decoup}), the crucial ingredient being that $Q$ commutes with the
non-linear algebra, thus factorising the averages:
\bea
Z [h,\j,A,\h ]
&=&\bigg\langle\exp\Big[ -\frac{1}{\pi}
\int ( h \tT + \j_a \tG^a + A_a \tU^a )\Big]
\Bigl\langle
    \exp\Big[ -\frac{1}{\pi} \int ( h T_Q + \hat{\h}Q)\Big]
\Bigr\rangle_{Q}\,\biggr\rangle\nonu
&=&\bigg\langle\exp\Big[ -\frac{1}{\pi} \int ( h \tT + \j_a \tG^a
+ A_a \tU^a )-\G [h,\hat{\h}]\Big] \biggr\rangle
\label{qdec}
\eea
where
\bea
&T_Q=\frac{3}{2c} Q\del Q,& \hat{\h}=\h-\frac{1}{3c} \j_aU^a.
\eea
The $Q$ integral can easily be expressed in terms of the Polyakov action:
\be
\G [h,\hat{\h}] = \frac{1}{48\pi} \G_{\mbox{Pol}}[h] - \frac{c}{6\pi}
\int\hat{\h} \frac{1}{\nablab}\hat{\h},\label{fermion}
\ee
where   $ \nablab=\bdel-h\del-\frac 1 2 \del h $ and
\be
\G_{\mbox{Pol}} [h]=\int\del^2h \frac{1}{\bdel}
\frac{1}{1-h\frac{\del}{\bdel}} \frac{1}{\del}\del^2h.
\ee
Using eqs. (\ref{qdec}) and (\ref{fermion}), we find
\bea
\exp\Big[ -\tGa[h,\j,A]\Big] &=&
%\exp\Big[ \frac{1}{48\pi}\G_{\mbox{Pol}}[h] \Big]
%\exp\Big[ -\frac{c}{6\pi}\int\Bigl(\h+\frac{\pi}{3c}\j_a\frac{\d}{\d
%A_a}\Bigr)\frac{1}{\nablab}
%\Bigl(\h+\frac{\pi}{3c}\j_b\frac{\d}{\d A_b}\Bigr) \Big] \nonu
%&&\qquad\exp \Big[ -\G [h,\j,A,\h ] \Big].\label{nll2}
\exp\Big[\G [h,\hat{\h}= \h+\frac{\pi}{3c}\j_b\frac{\d}{\d A_b}]\Big]
 \,\exp \Big[ -\G [h,\j,A,\h ] \Big].\label{nll2}
\eea
The double functional derivative in the exponential is well defined due to the
presence of the non-local operator $\nablab^{-1}$. This formula was checked
explicitly on the correlation functions using \cite{OPEdefs}.
%Actual computations are significantly simplified by rewriting $\nablab^{-1}$
%using $\bdel^{-1}_{x}\d^{(2)}(x-y)=\pi^{-1}(x-y)^{-1}$.
Introducing the Fourier transform of $\G$ w.r.t. $A$:
\be
\exp\Big[ -\G [h,\j,A,\h ]\Big] =\int [du]\exp\Big[ -\G [h,\j,u,\h
]+\frac{c}{3\pi}\int u^aA_a\Big],
\ee
eq. (\ref{nll2}) further reduces to
\bea
\exp\Big[ -\tGa[h,\j,A] \Big] &=&
\exp \Big[ \frac{1}{48\pi}\G_{\mbox{Pol}}[h] \Big]
\int [du] \exp\Big[ -\G [h,\j,u,\h ] \nonu
&&-\frac{c}{6\pi}\int\Bigl(\h+\j_au^a\Bigr) \frac{1}{\nablab}
\Bigl(\h+ \j_bu^b\Bigr)+\frac{c}{3\pi}\int u^aA_a\Big]  \quad
.\label{nll3}
\eea
As the lhs of eq. (\ref{nll3}) is $\h$-independent the rhs should also be.
We can integrate both sides over $\h$ with a measure chosen such
that the integral is equal to one:
\be
\exp\Big[   -\frac{1}{48\pi}\G_{\mbox{Pol}}[h]\Big]
\int [d\h] \exp\Big[ \frac{c}{6\pi}\int\Bigl(\h+\j_au^a\Bigr)
\frac{1}{\nablab}
\Bigl(\h+ \j_bu^b\Bigr)\Big] =1.
\ee
Combining this with eq. (\ref{nll3}), we obtain finally a very simple
expression for
$\tGa[h,\j,A]$ in terms of $\G [h,\j,u,\h ]$:
\be
\exp\Big[ -\tGa[h,\j,A] \Big] =
\int [d\h ]\exp\Big[ - \G [h,\j,u,\h ]\Big] . \label{nll4}
\ee
Introducing the generating functionals $W$
\bea
\exp \Big[ -W[t,g,u,q]\Big] &=& \int [dh][d\j][dA][d\h]\exp\Big[ -\G
[h,\j,A,\h ]\nonu
&&+\frac{1}{12\pi}\int (h\,t + 4 \j^a\,g_a -  4  A_a\, u^a -  4
\h\,q)\Big] \label{qquan}
\eea
and similarly for $\tW$ (without the $\h$-term),
one finds by combining eqs. (\ref{nll4},\ref{qquan}), an extremely simple
expression of the relation between the quantum theories of induced $N=3$
supergravities based on the linear and non-linear algebras:
\be
\tW[t,g,u]=W[t,g,u,q=0]. \label{W_{nl} in l}
\ee
Therefore, the two theories are related by a {\it quantum}
Hamiltonian reduction.

\section{$N=4$ Supergravity}
\setcounter{equation}{0}
Now we extend the method applied for $N=3$ to the case of
$N=4$. Again, there is a linear $N=4$ algebra and a
non-linear one, obtained \cite{goddard} by
decoupling 4 free fermions and a $U(1)$ current.
In the previous case we made use in the derivation of the
explicit form of the action induced by integrating out the fermions. In the
present case no explicit expression is available for the corresponding
quantity, but we will see that in fact it is not needed.

The $N=4$ superconformal algebra \cite{n4} is generated
by the energy-momentum tensor
$T$, 4 supercharges $G^a$, $a \in {\{}1,2,3,4{\}}$, an $so(4)$ affine Lie
algebra, $U^{ab} = -U^{ba}$, $a,b \in {\{}1,2,3,4{\}}$, 4 free fermions
$Q^a$ and a $U(1)$ current $P$.
The two $su(2)$- algebras  have levels $k_+$ and $k_-$.
The supercharges $G^a$ and the dimension $1/2$ fields $Q^a$ form two
$(2,2)$ representations of $SU(2) \otimes SU(2)$.
The central charge is given by :
\begin{equation}
c = {{6\,k_{+}\,k_{-}}\over {k_{+} + k_{-}}}\, .
\end{equation}
The OPEs are (we omit the OPES of $T$) :
\begin{eqnarray}
G^{a}\,G^{b} &=&
  {3c\over2} \delta ^{ab}[1] +
  [ -2\,U^{ab} + \zeta \,\vu ^{abcd} U^{cd}]
\nonu
U^{ab}\,U^{cd} &=&
  {k\over 2} \left( \delta^{ad}\,\delta ^{bc} - \delta^{ac}\,\delta ^{bd} -
                    \zeta \,\vu ^{abcd}
             \right) \,   [1] \nonu
&& +  [{\delta ^{bd}\,U^{ac} - \delta ^{bc}\,U^{ad} -
               \delta ^{ad}\,U^{bc} + \delta ^{ac}\,U^{bd}]}
\nonu
U^{ab}\,G^{c} &=&
       - \zeta \, \left( \delta ^{bc} \,[Q^{a}] - \delta ^{ac}\,[Q^{b}]
                  \right) + \vu ^{abcd}\,[Q^{d}]
 - (\delta ^{bc}\,[G^{a}] - \delta ^{ac}\,[G^{b}])
\nonu
Q^{a}\,G^{b} &=&
   \delta ^{ab}\,[P] - {1\over 2}\vu ^{abcd}\,[U^{cd}]
\nonu
Q^{a}\,U^{bc} &=&
  \delta ^{ac}\,[Q^{b}] - \delta ^{ab}\,[Q^{c}]
\nonu
P\,G^{a} &=& [Q^{a}]
\nonu
P\,P &=& -{k\over2}[1]
\nonu
Q^{a}\,Q^{b} &=&
  - {k\over 2}\delta ^{ab}[1]
   \label{eq:linalg}
\end{eqnarray}
where $k = k_{+} + k_{-}$ and $ \zeta= (k_{+} - k_{-})/ k$.

The induced action $\Gamma[h, \psi, A, b,\eta]$ is defined as in
(\ref{eq:indactL3}).  All the structure constants of the linear algebra
(\ref{eq:linalg}) depend only on the ratio $k_+/k_-$.  Apart from this
ratio, $k$ enters as a proportionality constant for {\it all} two-point
functions.  As a consequence, $\Gamma$
depends on that ratio in a non-trivial way, but its $k$-dependence is
simply an overall factor $k$.

Using the following definitions
\be
  t     = {12\pi \over c}  \DD{\Gamma}{h}, \hspace{1cm}
  g^a   = {3\pi \over c}   \DD{\Gamma}{\psi_a}, \hspace{1cm}
  u^{ab}= -{\pi \over k}   \DD{\Gamma}{A_{ab}}, \hspace{1cm}
  q^a   =-{2\pi \over k}  \DD{\Gamma}{\eta_a},  \hspace{1cm}
  p     =-{2\pi \over k}  \DD{\Gamma}{b}
\ee
and $ \gamma  = 6k / c$,
the Ward-identities are
\bea
\del^3 h &=& \nablab t
   - 2\left(\psi _a  \del + 3 \del \psi _a \right ) g^a
   + 2\,\gamma \,\del A_{ab}  u^{ab}
   +{\gamma \over 2}\,\left(\del \eta _a- \eta _a  \del \right) q^a
   + \gamma \,\del b \, p
   \nonu
\del^2 \psi _a &=& \nablab g^a
   -2\,A_{ab}  g^b
   - {1\over 2}\psi _a  t
   + {\gamma\over 4}\vu _{abcd}\,\eta _b  u^{cd}
   + {\gamma \over 4} \left(\psi _b \del + 2 \del \psi _b\right)
       (2u^{ab} - \zeta \vu _{abcd}\, u^{cd})\nonu
&& + {\gamma\over 4} \del b \, q^a
   + {\gamma\over 2} \zeta \,\del A_{ab}  q^b
   + {\gamma\over 4} \vu _{abcd}\,\del A_{bc}  q^d
   + {\gamma\over 4} \,\eta _a  p
   \nonu
\del A_{ab} &+& {{\zeta}\over 2} \vu _{abcd}\,\del A_{cd} =
\nablab u^{ab}
  -4\,A_{c[a}  u^{b]c}
          - {4\over {\gamma }}\psi _{[a}  g^{b]}
          + \eta _{[a}  q^{b]}
          - \zeta \left( \psi _{[a} \del + \del \psi_{[a} \right) q^{b]}
   \nonu
&&
   - {1\over 2}\vu _{abcd}\,\left( \psi _c \del + \del \psi_c \right) q^d
   \nonu
\eta_a &=& \nablab q^a
   -2\,A_{ab}  q^b - \psi _a  p +
   \vu _{abcd}\,\psi _b  u^{cd} \nonu
\del b &=& \nablab p
   -\left( \psi _a  \del + \del \psi _a \right) q^a \, .
   \label{eq:linearWard4}
\eea

The non-linear $N=4$ superconformal algebra has the same structure as
\ref{eq:linalg} but there is no $P$ and $Q^a$. The central charge is
related to the $su(2)$-levels by $\tc={3 (\tk + 2\tk_+ \tk_-) \over
2 + \tk}$.  We only give the $\tG \tG$ OPE explicitly:
\begin{eqnarray}
\tG^{a}\,\tG^{b} &=&
  {{4 \tk_{+} \tk_{-} }\over \tk+2} \, \delta ^{ab} \,[1]
   -  {2\tk \over \tk+2}  \, [\tU^{ab}]  +
       {\tk_+ - \tk_-\over \tk+2} \vu _{abcd}\,[\tU^{cd}]
    \nonu
&& +
[{2\tk \over \tk + 2 \tk_+ \tk_-} \,\delta ^{ab}\,\tT +
       {1 \over 4(\tk+2)}\,\vu _{acdg}\,\vu _{befg}
      (\tU^{cd} \tU^{ef} +  \tU^{ef} \tU^{cd})]
%\nonu
%\tU^{ab}\,\tU^{cd} &=&
% {1\over 2}\left( \tk \,
%            ( \delta^{ac}\,\delta ^{bd} - \delta ^{ad}\, \delta ^{bc}) -
%      {( \tk_{+} - \tk_{-}) \vu _{abcd}}
%     \right) [1] \nonu
%&&+
%  [\delta ^{bd}\,\tU^{ac} -
%      \delta ^{bc}\,\tU^{ad} -
%      \delta ^{ad}\,\tU^{bc} +
%      \delta ^{ac}\,\tU^{bd}]
%\nonu
%\tU^{ab}\,\tG^{c} &=&
%   - \delta ^{bc}\,[\tG^{a}] + \delta ^{ac}\,[\tG^{b} ]
\end{eqnarray}

To write down the Ward-identities in this case, we define
\be
  \ttt    =  {12\pi \over \tc}       \DD{\tGa}{h}, \hspace{1cm}
  \tg^a   = {(\tk+2) \pi\over 2\tk_+\tk_-} \DD{\tGa}{\psi_a}, \hspace{1cm}
  \tu^{ab}= -{\pi\over \tk}          \DD{\tGa}{A_{ab}},
\ee
and
\be
  \tgamma = {\tk (\tk+2) \over \tk_{+} \tk_{-}}, \hspace{1cm}
  \tkappa = {6 \tk \over \tc},\hspace{1cm}
  \tzeta = {\tk_{+}-\tk_{-} \over \tk}\, .
\ee
\bea
 \del^3 h&=&\nablab \ttt
   -{2\tkappa\over \tgamma} \left( \psi_a \del + 3 \del \psi_a\right) \tg^a
   + 2 \tkappa\,\del A_{ab} \tu^{ab}  \nonu
\del^2 \psi _a &=& \nablab \tg^a
   -2\,A_{ab}  \tg^b
   -{\tgamma\over 2\tkappa}\,\psi _a \ttt
   -{\tgamma\over 4 \tk (\tk+2)}
       \,\vu _{acdg}\, \vu _{befg}
       \, \psi _b \left(
           \left( \tU^{cd} \tU^{ef} \right)_{\it eff} +
           \left( \tU^{ef} \tU^{cd} \right)_{\it eff}
       \right)
       \nonu
&& + {\tgamma \tk \over 4(\tk+2)}\,(\psi_b \del +2\del \psi_b)
      ( 2\tu^{ab} - \tzeta \, \vu _{abcd} \tu^{cd} )
   \nonu
\del A_{ab} &+& {\tzeta\over 2}\,\vu _{abcd} \, \del A_{cd} =
\nablab \tu^{ab}
   -4 A_{c[a}  \tu^{b]c}
   -{4\over \tgamma} \psi _{[a}  \tg^{b]}
\eea

As in the previous section, we will use the results of \cite{goddard} on
the construction of a non-linear algebra by eliminating free fermion
fields.  In the present case it turns out that, at the same time, one can
also eliminate the $U(1)$-field $P$.  The new currents are
\begin{eqnarray}
\tT &=& T + {1\over k} P P +
   {1\over k}\del Q^{c} Q^{c}\nonu
\tG^{a} &=& G^{a} + {{2\over k}P Q^{a}} +
   \vu _{abcd}\,
    \left( {2\over 3k^2}\,Q^{b} Q^{c} Q^{d} +
      {1\over k}{Q^{b} \tU^{cd}} \right) \nonu
\tU^{ab} &=& U^{ab} - {2\over k}\,Q^{a} Q^{b}
 \label{eq:decompo}
\end{eqnarray}
and the constants in the algebras are related by
$\tk_\pm = k_\pm - 1$, and thus $ \tc = c-3$.
Again, we find agreement between the large $k$-limit of the Ward identities
putting $q^a$ and $p$ to zero, and solve $\eta^a$ and $\partial b$ from the
two last identities of (\ref{eq:linearWard4}). Note that the $b$-field
is present only as a derivative in eq. (\ref{eq:linearWard4}). Thus again the
suggestion presents itself that the non-linear action can be obtained by
putting some currents to zero.

We now set out to write the non-linear effective action in terms of the
effective action of the linear theory.
In analogy with the $N=3$ case, it would seem that, again, the operators of
the non-linear theory can be written as the difference of the operators of
the linear theory, and a realisation of the linear theory given by
the free fermions. In the present case, this simple linear combination
works for the integer spin currents $\tT$ and $\tU$, but not for $\tG$. A
second complication is that, due to the presence of a tri-linear term (in
$Q$) in the relation between $\tG$ and $G$, integrating out the
$Q$-fields is more involved. Nevertheless, we can still obtain
an explicit formula relating the effective actions.

There is a variety of ways to derive this relation, starting by
rewriting the decompositions of  (\ref{eq:decompo}) in different ways.
We will use the following form :
\begin{equation}
G^a + {1 \over k} \vu _{abcd} Q^b U^{cd} =
\tG^a - {2\over k} P Q^a + {4 \over {3 k^2}} \vu _{abcd} Q^b Q^c Q^d
\, .
\end{equation}
This leads immediately to
\footnote{%
Note that there are no normal ordering problems as the OPEs of the relevant
operators turn out to be non-singular (e.g. the term cubic in the $Q^a$
is an antisymmetric combination).}
\bea
\lefteqn{
  \Big\langle\exp\Big[
    -\frac{1}{\pi}  \int
         \Bigl( h T+\j_a G^a + A_{ab} U^{ab}  + b P+ \eta_a  Q^a
             + {1\over k} \vu_{abcd} \psi_a Q^b U^{cd}
         \Bigr)
   \Big]  \Big\rangle \, =
} \nonu
&&  \Big\langle\exp\Big[ -\frac{1}{\pi} \int
      \Bigl(
      \left(h \tT+\j_a \tG^a+ A_{ab} \tU^{ab} \right) \label{eq:startind4}\\
 &&\,\, +{1\over{ k}}\left(-h P^2-h \del Q^a Q_a- 2 \psi_a P Q^a
         + 2 A_{ab} Q^a Q^b + b P + \eta_a  Q^a \right)
        + {4 \over {3 k^2}} \vu _{abcd}\psi_a Q^b Q^c Q^d
\Bigr)\Big] \Big\rangle\, . \nonumber
\eea
Again the crucial step is that in the {\sl rhs}, the expectation value
factorizes: the average over $Q^a$ and $P$ can be computed separately,
since these fields commute with the non-linear SUSY-algebra.
This average is in fact closely related to the partition function
for the linear $N=4$ algebra with $k_+=k_-=1$ and $c=3$, up to the
renormalisation of some coefficients. We have
\bea
Z^{c=3}[h,\psi,A,b,\h]&=&\Big\langle\exp\Big[ -\frac{1}{\pi} \int
       \Bigl(-{h\over 2} \left(\hat{P}\hat{P} +\del \hat{Q}_a \hat{Q}^a
       \right)\label{eq:Zetdrie}\\
&&\quad -\psi_a \left(\hat{P}\hat{Q}^a+{1 \over 6}\vu _{abcd}
  \hat{Q}^b\hat{Q}^c\hat{Q}^d \right)
 +A_{ab}\hat{Q}^a \hat{Q}^b+b \hat{P} +\h_a \hat{Q}^a
\Bigr)\Big] \Big\rangle \nonumber
\eea
where the average value is over free fermions $\hat{Q}^a$ and a free
$U(1)$-current $\hat{P}$. These are normalised in a $k$-independent fashion
\bea
\hat{P}\hat{P}= -[1]&& \hat{Q}^a\hat{Q}^b=-\delta ^{ab}[1]
\eea
and the explicit form \cite{nis4cis3,n4} of the currents
making up the $c=3$
algebra has been used. The average can be represented as a functional
integral with measure
\be
[d\hat{Q}][d\hat{P}]\exp \Big[ -{1 \over {2\pi}}(\hat{P}{\bar{\del} \over
\del}\hat{P}+\hat{Q}^a\bar{\del}\hat{Q}_a)\Big] \quad. \label{eq:gewicht}
\ee
The (non-local) form of the free action for $\hat{P}$
follows from it's two-point
function: it is the usual (local) free scalar field action if one writes
$\hat{P}=\del \phi $.
The connection between the linear theory, the non-linear theory, and the
$c=3$ realisation is then
\bea
\lefteqn{
   \exp \Big{[} -{\pi\over k} \vu^{abcd} \psi_a
      \dd{ \eta_b}\dd{ A_{cd}}\Big{]}
      Z[\psi,A,\eta,b] =} \label{eq:ZtoZNL4}\\
&&\widetilde{Z}[\psi,A] \,\,
\exp \Big[ {{\pi^2}\over {3k^2}}(4+\sqrt{2k})  \vu^{abcd}
\psi_a \dd{ \eta_b}\dd{ \eta_c}\dd{ \eta_d}\Big{]}
Z^{c=3}[h,\psi,A,\eta\sqrt{k/2},b\sqrt{k/2}]  .\nonumber
\eea
Contrary to the $N=3$ case, where the Polyakov partition function was obtained
very explicitly, this connection is not particularly useful, but the
representation (\ref{eq:gewicht}) of $Z^{c=3}$ as a functional integral
can be used effectively. Indeed, when
we take the Fourier transform of eq. \ref{eq:ZtoZNL4}, i.e. we integrate
(\ref{eq:ZtoZNL4}) with
\be \int [d\,h][d\,\psi][d\,A][d\,b][d\,\h]\exp \Big[ {1 \over \pi }
    \int\Bigl( h\,t +\j_a\,g^a + A_{ab}\, u^{ab}  + b\, p +
\h_a\,q^a\Bigr)\Big] \quad ,
\ee
we obtain using eqs. \ref{eq:Zetdrie} and \ref{eq:gewicht}
\footnote{
The effective action $W$ is defined by the Fourier transform of $Z$, and
similarly for $\tW$}.
\bea
\lefteqn{
\exp \Big[ -W[t,\,g^a-{1\over k}\vu _{abcd}q^bu^{cd},\,u,\,p,\,q] \Big] =
\exp \Big[ -{1\over{\pi k}}(p{\bar{\del}\over \del}p+
                     q^a\bar{\del}q_a) \Big]}\nonu
&&\exp \Big[-\tW[t+{1\over k}\left(p^2+\del q\,q \right),\,
g^a+{2\over k}p\,q^a-{4\over {3k^2}} \vu _{abcd}q^bq^cq^d,\!
u^{ab}-{2\over k}q^aq^b] \Big] \nonu
&&
\eea
giving the concise relation
\bea
\tW[t,\,g^a,\,u^{ab}]
&+&{1\over{\pi k}}(p{\bar{\del}\over \del}p+q^a\bar{\del}q_a)\nonu
&=&W[t-{1\over k}\left(p^2+\del q^a\,q_a \right),\nonu
&&\qquad  g^a-{2\over k}pq^a-{1\over k}\vu _{abcd}q^bu^{cd}
                 -{2\over {3k^2}}\vu _{abcd}q^bq^cq^d,
          u^{ab}+{2\over k}q^aq^b,\,p,\,q^a] \nonu
\eea
Putting the free $p$ and $q^a$-currents equal to zero, one obtains the
equality of effective actions:
\be
\tW[t,g^a,u^{ab}]=W[t,g^a,u^{ab},p=0,q^a=0]. \label{W_nl in l4}
\ee
\section{Discussion}
\setcounter{equation}{0}

We take for
granted that the linear theory is given by simple renormalisations of the
'classical' theory, as described in the introduction. Then eqs. (\ref{W_{nl}
in l}) and (\ref{W_nl in l4}) immediately transfer this property to the
non-linear theory.
Moreover, since the 'classical' parts are equal also (as implied by the
$c \rightarrow \infty$ limit of the Ward identities) the renormalisation
factors for both theories are the
same (for couplings as well as for fields) if one takes into account the
shifts in the values for the central extensions $c$, $ k_+$ and $k_-$.
This fact can be confirmed by looking at explicit calculations of these
renormalisation factors. For $N=3$, a semiclassical approximation to the
non-linear renormalisation factors was set up in \cite{gustav}. This
calculation was amended in \cite{zfactors}, which also contains an all-order
calculation of these factors. On the other hand, \cite{zfactors} also contains
a semiclassical derivation of the factors for the {\it linear} algebras, and
the $N=4$ factors as well.
The results are:
\[
\begin{array}{llclcl}
               &\mbox{\hspace*{1cm}}&\mbox{non-linear all-order}
 &\hspace*{1cm}&\mbox{linear semiclassical}&\\
\\
\fbox{$N = 3$} &                    &Z_\Gamma = {2\tc+1 \over 2} -3
 &\hspace*{1cm}&Z_\Gamma  = c-3\\
               &                    &Z_h = {2\tc+1\over 2\tc-5}
 &             &Z_h =  {c \over c-3}\\
\\
\fbox{$N = 4$} &\mbox{\hspace*{1cm}}&Z_\Gamma  = {\tc+3}
 &             &Z_\Gamma = c\\
              &                     &Z_h = 1
 &             & Z_h = 1
\end{array}
\]
and the other field renormalisation factors ($Z_\j, Z_u, Z_q, Z_p$)
for the effective action are the same.
Clearly, these results coincide if one takes into account that
$c = \tc + \half $ ({\sl resp. }$\tc + 3$).

The remarkable property that the values of the central extensions
of the Virasoro and affine algebras are related by
$c=6 k +n$ with $n \in {\bf Z}$, is shared by a number of other
non-linear superconformal and quasisuperconformal algebras.
These contain only one dimension two field, a number of (bosonic and
fermionic) dimension 3/2 fields, and an affine superalgebra
by which we will identify them.
There are the $osp(m|2n)$ cases with $|m-2n-3|\leq 1$
(with $m=3,4$ and $n=0$ treated in this paper, and $m=2,n=0$ the
ordinary $N=2$ superalgebra)
and the $u(n|m)$ cases with $|n-m-2| \leq 1$  from the series in
 \cite{filipzbig},
and further the $osp(n|2m)\oplus sl_2$ algebras  with $|n-2m+3|\leq 1$
from  \cite{fradlin}.
These same algebras arise also (among others) by quantum
Drinfeld-Sokolov reduction from the list
\cite{zfactors} where there are no corrections to the coupling beyond
one loop. This is reminiscent of $N=2$
supergravity, and consequently also of supersymmetry non-renormalisation
theorems\cite{nonrenor}, but the evidence is not conclusive.
It would be interesting to investigate whether one can extend the
analysis of the present paper in this direction.


\newpage
\frenchspacing
\begin{thebibliography}{11}
\bibitem{pbks}
  P. Bouwknegt and K. Schoutens, preprint CERN-TH.6583/92 and
ITP-SB-92-23, to be published in Phys. Rep.
\bibitem{zamo}
  A.B. Zamolodchikov, Theor.Math.Phys. {\bf 63} (1985) 1205
\bibitem{polyakov}
  A. M. Polyakov,  Mod. Phys. Lett. {\bf A2} (1987) 893
\bibitem{KPZ}
 V. G. Knizhnik, A. M. Polyakov and A. B. Zamolodchikov,
 Mod. Phys. Lett {\bf A3} (1988) 819
\bibitem{Zfac methods}
 V.G.Knizhnik and A. B. Zamolodchikov, \np {\bf B247} (1984) 83\\
 Al. B. Zamolodchikov, preprint ITEP 84-89 (1989)\\
 K. A. Meissner and J. Pawe{\sll}czyk, Mod. Phys. Lett {\bf A5} (1990) 763\\
 K. Schoutens, A. Sevrin and P. van Nieuwenhuizen, proc. Stony Brook
 conference {\it Strings and Symmetries 1991}, World Scientific (1992)
\bibitem{ruud}
  A. Sevrin, R. Siebelink and W. Troost, in preparation.
\bibitem{ssvnb}
  H. Ooguri, K. Schoutens, A. Sevrin and P. van Nieuwenhuizen,
  Comm. Math. Phys. {\bf 145} (1992) 515
\bibitem{dbg}
  J. de Boer and J. Goeree, Utrecht preprint THU-92/33
\bibitem{zfactors}
  A. Sevrin, K. Thielemans and W. Troost,
  LBL-33738, UCB-PTH-93/06, KUL-TF-93/09
\bibitem{drisok} V. G. Drinfeld and V. V. Sokolov, J. Sov. Math. {\bf 30}
 (1984) 1975
\bibitem{adem}
M. Ademollo et al., \pl {\bf 62B} (1976) 105; \np {\bf B111} (1976) 77;
\np {\bf B114} (1976) 297
\bibitem{n4}
 K. Schoutens, \np {\bf B295[FS21]} (1988) 634 \\
 A. Sevrin, W. Troost and A. Van Proeyen, \pl {\bf B208} (1988) 447
\bibitem{bershkniz} M. Bershadsky, \pl {\bf B174} (1986) 285\\
  V. G. Knizhnik, Theor. Math. Phys. {\bf 66} (1986) 68
\bibitem{goddard} P. Goddard and A. Schwimmer, \pl {\bf 214B} (1988) 209
\bibitem{gustav} G.W. Delius, M.T. Grisaru and P. van Nieuwenhuizen,
preprint CERN-TH.6458/92
\bibitem{nis4cis3}
  K. Schoutens, \pl {\bf B194} (1987) 75
\bibitem{OPEdefs}
  K. Thielemans, Int. J. Mod. Phys. {\bf C2} (1991) 787\\
  K. Thielemans, Proc. of the Second Int. Workshop on
Software Engineering, Artificial Intelligence and Expert Systems in High
Energy and Nuclear Physics (1992), World Scientific, p. 709
\bibitem{filipzbig} F. Defever, Z. Hasiewicz and W. Troost, \pl {\bf B273}
 (1991) 51
\bibitem{fradlin} E. S. Fradkin and V. Linetsky, \pl {\bf B291} (1992) 71
\bibitem{nonrenor}M. T. Grisaru, W. Siegel and M. Ro\v{c}ek, \np {\bf B159}
(1979) 429; {\it Superspace}, S. J. Gates, M. T. Grisaru, W. Siegel and M.
Ro\v{c}ek, Benjamin/Cummings pub. comp. 1983, p358.

\end{thebibliography}
\end{document}